\documentstyle[aps,epsf]{revtex}

\twocolumn

\begin{document}

\title{Parametric Oscillatory Instability in Fabry-Perot (FP) Interferometer}
\author{V. B. Braginsky, S. E. Strigin and S. P. Vyatchanin}
\address{Physics Faculty, Moscow State University,
    Moscow 119899, Russia \\
    e-mail: vyat@hbar.phys.msu.su}
\date{\today}
\maketitle
\begin{abstract}
We present an approximate analysis of a nonlinear effect of
parametric oscillatory instability in FP interferometer. The basis
for this effect is the excitation of the additional (Stokes)
optical mode with frequency $\omega_1$ and of the mirror's elastic mode
with frequency $\omega_m$ when the optical energy stored in the main
FP resonator  mode with frequency $\omega_0$ exceeds the certain
threshold and the frequencies are related as $\omega_0\simeq
\omega_1+\omega_m$. This effect is undesirable in 
laser gravitational wave antennae because it may create a specific
upper limit for the value of energy stored in FP resonator. In
order to avoid it the detailed analysis of the mirror's elastic
modes and FP resonator optical modes structure is necessary.
\end{abstract}

\section{Introduction}

The full scale terrestrial gravitational wave antennae are in
process of assembling and tuning at present. One of these antennae
(LIGO-I project) sensitivity expressed in terms of the metric
perturbation amplitude is projected to achieve  soon the level of
$h\simeq 1\times 10^{-21}$ \cite{abr1,abr2}. In 2008 the
projected level of sensitivity has to be not less than $h\simeq
1\times 10^{-22}$ \cite{amaldi}. This value is scheduled to
achieve by substantial improvement of the test masses (mirrors in
the big FP resonator) isolation from different sources of noises
and by increasing the optical readout system sensitivity. This
increase is expected to be obtained by rising the value of
optical energy ${\cal E}_0$ stored in the FP resonator optical
mode: ${\cal E}_0 > 30$~J (it corresponds to the circulating power
$W$ bigger than $1$~megaWatt). So high values of ${\cal E}_0$ and
$W$ may be a source of the nonlinear effects which will prevent
from reaching the projected sensitivity of $h\sim 1\times
10^{-22}$. Authors of this article already described two such
effects: photo-thermal shot noise \cite{bgv} (the random
absorption of optical photons in the surface layer of the mirror
causes the fluctuating of mirror surface due to nonzero
coefficient of thermal expansion) and photo-refractive shot noise
\cite{bgvrefr} (the same random absorption of optical photons
causes the fluctuations of the reflected wave phase due to the
dependence of refraction index on temperature). In this paper we
analyze undesirable effect of parametric instability
--- another "trap" of pure dynamical nonlinear origin which (being
ignored) may cause very substantial decrease of the antennae
sensitivity
 and even may make the antenna unable to work properly.

It is appropriate to remind that nonlinear coupling of elastic and
light waves in continuous media produces Mandelstam-Brillouin
scattering. It is a classical  parametric effect, however, it is
often explained in terms of quantum physics: one quantum
$\hbar\omega_0$ of main optical wave transforms into two, {\it i.
e.} $\hbar \omega_1$ in the additional optical wave (Stokes wave:
$\omega_1<\omega_0$) and $\hbar \omega_m$ in the elastic wave so
that $\omega_0=\omega_1 +\omega_m$
(it is Manley-Rowe condition for parametric process).
The irradiation into the anti-Stokes wave is also possible
($\omega_1=\omega_0 +\omega_m$), however, in this case the part of
energy is taken from the elastic wave. The physical "mechanism" of
this coupling is the dependence of refractive index on density
which is modulated by elastic waves. If the main wave power is
large enough the stimulated scattering will take place, the
amplitudes of elastic and Stokes waves will increase
substantially. The physical description is the following: the flux
of energy into these waves is so large that before being
irradiated from the volume of interaction, the oscillations with
frequencies $\omega_1$ and $\omega_m$ stimulate each other
substantially increasing the power taken from the main wave. Note
that stimulated scattering causes irradiation only into Stokes
wave because the additional
energy pump into elastic wave must take place 
for radiation into anti-Stokes wave.

In gravitational wave antennae  elastic
oscillations in FP resonator mirrors 
will interact with optical ones being coupled
parametrically due to the boundary conditions on  one hand, and
due to the ponderomotive force on the other hand. Two optical
modes may play roles of the main and Stokes waves. High quality
factors of these modes and of the elastic one will increase the
effectiveness of the interaction between them and may give birth
to the parametric oscillatory instability which is similar to
stimulated Mandelstam-Brillouin effect \cite{french}. This
instability may create a specific upper limit for the value of
energy ${\cal E}_0$.

It is worth to note that this effect of parametric instability is
a particular case of the more general phenomenon related to the
dynamical back action of parametric  displacement  meter on 
mechanical oscillator or
free mass. This dynamical back action was
analyzed and observed more than 30 years ago \cite{BM1,BM2}.
Usually parametric meter consists of e.m.
resonator with high quality factor $Q$ (radiofrequency, microwave 
or optical ones) and high
frequency stability pumping self-sustained oscillator. The
displacement of the resonator movable element modulates its
eigenfrequency which in its turn produces the modulation of the
e.m. oscillations amplitude (it can produce also phase or output
power modulations). If the experimentalist attaches a probe mass
to the movable element of the meter he is inevitably confronted
with the effect of dynamical back action, the ponderomotive force
produces a rigidity and due to finite e.m. relaxation time --- a
mechanical friction. Both these values may be positive and
negative ones. In the case when the negative friction is
sufficiently high the behavior of mechanical oscillator and meter
becomes oscillatory unstable. This effect was observed and
explained for the case when value of mechanical frequency
$\omega_m$ was substantially smaller than the bandwidth of e.m.
resonator \cite{BM1,BM2}. In this article we analyze the
parametric oscillatory instability in two optical modes of FP
resonator and elastic mode of the mirror. In this case the value
of $\omega_m$ is much larger than the bandwidth of the optical
modes.

In section II we present the analysis of this effect for
simplified one-dimensional model which permits to obtain
approximate estimates for the instability conditions. In section
III we present considerations and  preliminary estimates for
nonsimplified three-dimensional model, and in section \ref{sec3} ---
the program of necessary mode numerical analysis.

\section{Simplified one-dimensional model}\label{sec1}

\begin{figure}
\noindent
\unitlength0.11mm
\linethickness{0.5pt}
\begin{picture}(790,543)
\put(120,275){\framebox(40,240)[]{}}
\put(480,275){\framebox(40,240)[]{}}
\put(20,475){\vector(1,0){100}}
\put(120,315){\vector(-1,0){100}}
\put(280,315){\vector(-1,0){120}}
\put(360,475){\vector(1,0){120}}
\put(480,315){\vector(-1,0){120}}
\put(160,475){\vector(1,0){120}}
\put(520,395){\line(1,2){40}}
\put(560,475){\line(1,-2){80}}
\put(640,315){\line(1,2){40}}
\put(680,515){\line(0,-1){240}}
\put(680,475){\line(1,1){40}}
\put(680,435){\line(1,1){40}}
\put(680,395){\line(1,1){40}}
\put(680,315){\line(1,1){40}}
\put(680,275){\line(1,1){40}}
\put(680,355){\line(1,1){40}}
\put(540,275){$x$}
\put(480,245){\vector(1,0){80}}
\put(480,265){\line(0,-1){40}}
\put(0,220){\framebox(741,321)[]{}}
\put(540,360){$m$}
\put(140,275){\line(0,-1){35}}
\put(100,240){\line(1,0){80}}
\put(300,80){\vector(1,0){420}}
\put(660,80){\line(0,1){20}}
\put(640,40){$\omega_0$}
\put(660,80){\line(0,1){20}}
\bezier{80}(580,80)(620,80)(620,120)
\bezier{60}(720,80)(700,80)(700,120)
\bezier{100}(620,120)(620,180)(660,180)
\bezier{100}(660,180)(700,180)(700,120)
\put(40,80){\line(1,0){280}}
\put(20,100){\line(0,-1){40}}
\put(20,80){\line(1,0){20}}
\put(0,0){\framebox(740,220)[]{}}
\put(40,500){{\bf a)}}
\put(40,180){{\bf b)}}
\put(180,100){\line(0,-1){20}}
\put(120,120){\vector(1,0){40}}
\put(240,120){\vector(-1,0){40}}
\put(160,40){$\omega_m$}
\bezier{25}(160,81)(170,85)(170,100)
\bezier{64}(170,101)(173,158)(180,158)
\bezier{68}(190,100)(193,158)(183,158)
\bezier{26}(190,100)(193,81)(200,80)
\put(220,140){$2\delta_m$}
\put(500,80){\line(0,1){20}}
\put(480,40){$\omega_1$}
\bezier{80}(420,80)(460,80)(460,120)
\put(420,140){\vector(1,0){40}}
\put(580,140){\vector(-1,0){40}}
\bezier{60}(560,80)(540,80)(540,120)
\bezier{100}(460,120)(460,180)(500,180)
\bezier{100}(500,180)(540,180)(540,120)
\put(540,160){$2\delta_1$}
\end{picture}
\\[2mm]
\caption{Scheme of FP resonator with movable mirror (a) and
    frequency diagram (b).}\label{fig1}
\end{figure}
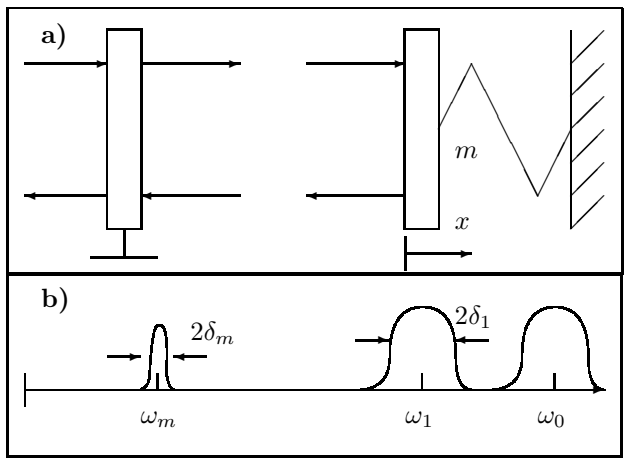

For approximate estimates we present in this section the
simplified model analysis where we assume that:
\begin{itemize}
\item
    The mechanical oscillator (model of mirror) is a
    lumped one with single mechanical degree of freedom (eigenfrequency
    $\omega_m$ and quality factor $Q_m=\omega_m /2\delta_m$).
\item
    This oscillator mass $m$ is the FP resonator right mirror
    (see fig.~\ref{fig1}) having ideal reflectivity
    and the value of $m$ is  of the order of the total mirror's mass.
\item 
    The left mirror (through which FP resonator is pumped) has an infinite
    mass, no optical losses and finite transmittance
    $T=2\pi L/(\lambda_0 Q_{opt})$ ($\lambda_0$ is the optical wavelength,
    $Q_{opt}$ is the quality factor, $L$ is the distance between the
    mirrors).
\item
    We take into account only the main mode with frequency  $\omega_0$
    and relaxation  rate $\delta_0=\omega_0/2Q_0$  and
    Stokes mode with $\omega_1$ and $\delta_1=\omega_0/2Q_1$
    correspondingly ($Q_0$ and $Q_1$ are the quality factors),
    $\omega_0-\omega_1 \simeq \omega_m$.
\item
    Laser is pumping only the main mode which stored energy ${\cal E}_0$
    is assumed to be a constant one (approximation of
    constant field).
\end{itemize}
It is possible to calculate at what level of energy ${\cal E}_0$
the  Stokes mode and mechanical oscillator becomes unstable.
The origin of this instability can be described qualitatively in
the following way: small mechanical oscillations with the
resonance frequency $\omega_m$ modulate the distance $L$ that
causes the exitation of optical fields with frequencies 
$\omega_0 \pm \omega_m$. Therefore, the Stokes mode amplitude will rise
linearly in time if time interval is shorter than $\delta_1^{-1}$.
The presence of two optical fields with frequencies $\omega_0$ and
$\omega_1$ will produce the component of ponderomotive force
(which is proportional to square of sum field) on difference
frequency $\omega_0-\omega_1$. Thus this force will increase the
initially small amplitude of mechanical oscillations. In other
words, we have to use two equations for  Stokes mode
and mechanical oscillator and find the conditions when this
"feedback" prevails the damping which exists due to the finite values
of $Q_m$ and $Q_1$. Below we present only the scheme of
calculations (see details in Appendix \ref{app1}).

We write down the field components of optical modes
and the displacement $x$ of mechanical oscillator in rotating wave
approximation:
\begin{eqnarray*}
E_0 &=& A_0[D_0e^{-i\omega_0 t}+D_0^*e^{i\omega_0 t}],\\
E_1 &=& A_1[D_1e^{-i\omega_1 t}+D_1^*e^{i\omega_1 t}], \\
x &=& Xe^{-i\omega_m t} + X^* e^{i\omega_m t},
\end{eqnarray*}
where $D_0$ and $D_1$ are the slowly changing complex amplitudes
of the main and Stokes modes correspondingly and $X$ is the slowly
changing complex amplitude of mechanical displacement. Normalizing
constants $A_0,\ A_1$ are chosen so that energies ${\cal
E}_{0,\,1}$ stored in each mode are equal to ${\cal E}_{0,\,1}=
\omega_{0,\, 1}^2|D_{0,\, 1}|^2/2$. Then it is easy to obtain the
equations for slowly changing amplitudes:
\begin{eqnarray}
\label{Q1}
\partial_t D_1 +\delta_1 D_1  &=&
    \frac{iX^*\,D_0 \omega_0 }
        {L }\, e^{-i\Delta\omega t},\\
\label{X}
\partial_t X +\delta_m X &=&
    \frac{i\, D_0D_1^*\omega_0\omega_1}{m\omega_m L }\,
    \,e^{-i\Delta\omega t},
\end{eqnarray}
where $\Delta\omega = \omega_0-\omega_1-\omega_m$ is the possible
detuning. Remind that we assume $D_0$ as a constant.

One can find the solutions of (\ref{Q1}, \ref{X}) in the following
form $D_1(t) = D_1 e^{(\lambda - \Delta \omega/2) t}$, $X^*(t) =
X^* e^{(\lambda + \Delta \omega/2) t}$ and write down the
characteristic equation. The parametric oscillatory instability
will appear if real part of one of the characteristic equation roots 
is positive.

In LIGO design the values of $\delta_0$ and $\delta_1$ are of the
order of the bandwidth the gravitational burst spectrum is
expected to lie in, i.e.
    {\bf $\simeq 2\pi\times 100\ {\rm s}^{-1} $}.
On the other hand many efforts were made to reduce the value of
$\delta_m$ to the lowest possible level and thus to decrease the
threshold of sensitivity caused by Brownian noise. In existing
today fused silica mirrors $Q_m\simeq 10^6 - 2\times 10^7$ and
even for $\omega_m =10^7\ {\rm s}^{-1}$ the value of $\delta_m \le
10\ {\rm s}^{-1}$. Thus we can assume that $\delta_m \ll \delta_1$ and 
obtain the instability condition in simple form:
\begin{eqnarray}
\label{CI}
\frac{{\cal R}_0}{\left(1+\frac{\Delta \omega^2}{\delta_1^2}\right) } & > & 1,\\
\label{R0}
{\cal R}_0  =  \frac{{\cal E}_0 }{2mL^2\omega_m^2}\,
    \frac{\omega_1\omega_m}{\delta_1\delta_m}&=&
    \frac{{2\cal E}_0 Q_1Q_m}{mL^2\omega_m^2}\, .
\end{eqnarray}

For estimates we assume parameters corresponding to LIGO-II to be:

\begin{equation}
\label{param}
\begin{array}{lcllcl}
\omega_m &=& 2\times 10^5 \ {\rm sec}^{-1},&
    \delta_m &=&5\times 10^{-3} \ {\rm sec}^{-1},\\
\delta_1 &=& 6\times 10^2\ {\rm sec}^{-1}, &
    \omega_1& \simeq & 2\times 10^{15}\ {\rm sec}^{-1},\\
{\cal E}_0 &\simeq& 3\times 10^8\ {\rm erg},& L&=&4\times 10^5\ {\rm cm}\\
m &=& 10^4\ {\rm g},&  &&
\end{array}
\end{equation}

The mechanical frequency $\omega_m$ is about the frequency of the
lowest mirror elastic (longitudinal or drum) mode and it has the
same order as the intermodal interval $\sim \pi c/L \simeq 2\times
10^5\ {\rm sec}^{-1}$ between optical modes of FP resonator. The
mechanical relaxation rate $\delta_m$ corresponds to the loss
angle $\phi\simeq 5\times10^{-8}$ (quality factor $Q_m\simeq
2\times 10^7$) for fused silica. The value of energy ${\cal E}_0$
corresponds to the value of circulating power about $W\simeq
c{\cal E}_0/2L \simeq 10^{13}\ \mbox{erg/s}=10^6$~Watt.

For these parameters  we have obtained the estimate of coefficient 
${\cal R}_0$  for the resonance case ($|\Delta\omega|\ll \delta_1
$):

$$ {\cal R}_0 \simeq 300 \gg 1 $$

It means that the critical value of stored energy ${\cal E}_0$ for
the instability initiation will be 300 times smaller than the
planned value $\simeq 3\times 10^{8}\ {\rm erg}=30$~J \footnote{
    It is worth to note that if sapphire is chosen then due to
   the larger value of $Q_m$ the factor ${\cal R}_0$ will be even bigger:
    $
    {\cal R}_0 \simeq 5\times10^{3}.
    $
    It is another argument which is not in favor of this material for mirrors.}.

For nonresonance case and planned value of ${\cal E}_0$
the "borders" of detuning $\Delta \omega_{crit}$ within the system
is unstable, are relatively large: 
$\Delta \omega_{crit}= \delta_1\sqrt{R_0}\simeq 1.7\times 10^4\ {\rm sec}^{-1}$.

\section{ Considerations of Three-Dimensional Modes Analysis}\label{sec2}

The numerical estimates for the values of factor ${\cal R}_0$ and
detuning $\Delta \omega$ obtained in the preceding section have to
be regarded as some kind of warning about the reality of the
undesirable parametric instability effect. In the simplified
analysis we have ignored the nonuniform distribution of optical
fields and of mechanical displacements over the mirror's surface.
It is evident that  more accurate analysis has to be done. Below
we present several considerations about further necessary
analysis.

\subsection{The frequency range of "dangerous" optical and elastic modes}

The values of the mirror's radius $R$ and thickness $H$ for
LIGO-II are not yet finally defined. Due to the necessity to
decrease the level of thermoelastic and thermorefractive noises
\cite{bgv,bgvrefr,kip,thorneft} the size of the light spot on the
mirror's surface is likely to be substantially larger than in
LIGO-I and the light density distribution in the spot is not
likely to be a gaussian one (to evade  substantial diffractional losses)
\cite{thorneft}. Thus the presented below estimates for gaussian optical 
modes may be
regarded only as the first approximation in which the use of
analytical calculations is still possible.

The resonance conditions $|\omega_0 -\omega_1 -\omega_m|<
\delta_1$ may be obtained with a relatively high probability for
many optical Stokes and mirror elastic modes combinations. If we
assume the main optical mode to be gaussian one with waist radius
$w_0$  of the caustic (the optical field amplitude distribution in
the middle between the mirrors is $\sim e^{-r^2/w_0^2}$)), and if
we assume also that the Stokes mode may be described by
generalized Laguerre functions (Gauss-Laguerre beams) then the set
of frequency distances $\Delta \omega_{\rm opt}$ between the main
and Stokes modes is determined by three integer numbers:
\begin{equation}
\label{DeltaOpt}
\Delta \omega_{\rm optic} \simeq \frac{\pi c}{L}\left(K +
    \frac{2(2N+M)}{\pi}\, \arctan\frac{L\lambda_0}{2\pi w_0^2}\right),
\end{equation}
where $\lambda_0$ is the wave length, $K=0\pm 1,\, \pm 2\dots $ is
the longitudinal index, $N=0,\, 1,\, 2\dots $, and $M=0,\, 1,\,
2\dots $ are the radial and angular indices.

For $w_0\simeq 5.9$~cm the beam radius on the mirror's surface is
equal to $w\simeq 6$~cm, corresponding to the level of
diffractional losses about 20 ppm for mirror radius of $R=14$~cm.
In this case the equation (\ref{DeltaOpt}) has the following form:
\begin{equation}
\label{Deltaomegaopt}
\Delta \omega_{\rm optic} \simeq \left(2.4\, K +
     0.56\, N + 0.28\, M  \right)\times 10^5 \,
    \mbox{s}^{-1}.
\end{equation}
We see that the distance  between optical modes  is not so large,
i.e. $\simeq 3\times 10^4\, {\rm s^{-1} }$. In units of optical
modes bandwidth $2\delta_1\simeq 10^3\, {\rm s^{-1} }$  it is
about $ 3\times 10^4/2\delta_1\simeq 30$. Thus assuming that the
value of elastic mode frequency can be an arbitrary one we can
roughly estimate the  probability that the resonance condition is
fulfilled as $\sim 1/30$.

The order of the distance $\Delta \omega_{m0}$ between the
frequencies for the first several elastic modes is about $\Delta
\omega_{m0} \simeq \pi v_s/d\simeq 2\times 10^5$~s$^{-1}$ ($d$ is
the dimension of the mirror and $v_s$ is the sound velocity). It
is about one order larger than the distance between the optical
modes. However, for higher frequencies $\omega_m$ these intervals
become smaller and can be estimated by formula $$ \Delta\omega_m
\simeq \frac{\pi \Delta\omega_{m0}^3}{2\omega_m^2} $$ Even for
$\omega_m\simeq 6\times 10^5\ \mbox{s}^{-1}$ the intervals between
the elastic and  optical modes become equal to each other and has
value about $\simeq 3 \times 10^{4}\,
\mbox{s}^{-1}$. And for $\omega_m \sim 10^7\ {\rm s}^{-1}$ the
distances between elastic modes become of the order of optical
bandwidth $2\delta_1$. Therefore, the resonance condition for these
frequencies is practically always fulfilled.

On the other hand according to (\ref{R0}) the factor ${\cal R}_0$
decreases  for higher elastic frequencies $\omega_m$. In addition
the loss angle in fused silica usually slightly increases for higher
frequencies \cite{mason,numata}. Assuming that the upper value
 $\omega_m\simeq 2\times 10^6\ {\rm s}^{-1}$,
$Q_m \simeq 3\times 10^{6}$ and other parameters correspond to
(\ref{param})  we obtain $ {\cal R}_0\simeq 1 $. Therefore, the
elastic modes which "deserve" accurate calculations lie within
the range between several tens and several hundreds kiloHerz. The
total number of these  modes is about  several hundreds.

\subsection{The Matching between the Mechanical Displacements
    and Light Density Distributions}

The simplified model described in section \ref{sec1} is
approximately valid for the uniform over all the mirror's surface
distribution of optical field density and pure longitudinal
elastic mode. The equations for this model can be extended for any
distributions of mechanical displacements in the chosen elastic
mode and for any distribution of the light field density in chosen
optical modes. This extension can be done by adding a
dimensionless factor $\Lambda$ in (\ref{CI}):
\begin{eqnarray}
\label{CIL}
\frac{{\cal R}_0 \Lambda}{
    \left(1+\frac{\delta \omega^2}{\delta_1^2}\right) } & > & 1,\\
\label{Lambda}
\Lambda  =  \frac{V\left(\int f_0(\vec r_\bot)\, f_1(\vec r_\bot)\,
        u_{z}\, d\vec r_\bot\right)^2}{
    \int |f_0|^2 d\vec r_\bot\, \int |f_1|^2 d\vec r_\bot \,
        \int |\vec u|^2 d V }.
\end{eqnarray}
Here $f_0$ and $f_1$ are the functions of the distributions over
the  mirror's surface of the optical fields in the main and Stokes
optical modes correspondingly, vector  $\vec u$ is the spatial
vector of displacements in elastic mode, $u_z$ is the component of
$\vec u$, normal to the mirror's surface,  $\int d\vec r_\bot$
corresponds to the integration over the mirror's surface and $\int
dV$
--- over the mirror's volume $V$.

It is necessary to know the functions $f_0,\, f_1,\ \vec u$ in
order to calculate the factors $\Lambda$ for different mode
combinations. But to the best of our knowledge there is no
analytical form for $\vec u$ in the case of cylinder with free
boundary conditions \cite{mason}. Our approximate estimates
show that there is substantial number of modes combinations for
which factor $\Lambda$  is large enough to satisfy the condition
(\ref{CIL}) for parameters (\ref{param}). We do not present the
details of these estimates here because they are rather rough. It
is evident that a complete numerical analysis which includes
nongaussian distributions of optical fields is necessary to
do.

\section{Conclusion}\label{sec3}

The  simplified model analysis of parametric oscillatory instability
and considerations about the real model presented above may be
regarded only as the first step along the route to obtain a
guarantee to evade this nondesirable effect. Summing up we may
formulate several recommendations for the next steps:

\begin{enumerate}
\item
Due to the finite size of the  mirror and to the use of the
nongaussian distribution of light density  we think that the
accurate numerical analysis of different optical and elastic mode
combinations (candidates for the parametric instability) is
inevitably necessary. This problem (numerical
calculations for the elastic modes) has been already solved partialy
\cite{raab}.

\item
In the same time the numerical analysis may not give an absolute
guarantee because the fused silica pins and fibers will be
attached to the mirror. This attachment will change the elastic
modes frequency values (and may be also the distribution). In
addition the unknown Young modulus and fused silica density
inhomogeneity will limit the numerical analysis accuracy. Thus the
direct measurements for several hundreds of probe mass elastic
modes eigenfrequencies values and quality factors are also
necessary.

\item
When more "dangerous" candidates of elastic and Stokes modes will
be known, their undesirable influence can be depressed. For
example it can be done  by small the change of mirror's shape.


\item
It is also reasonable to perform direct tests of the optical field
behavior with smooth increase of the input optical power: it will
be possible to register the appearence of the photons  at the Stokes modes
and the rise of the $Q_m$ in the corresponding elastic mode until
the power $W$ in the main optical mode is below the critical
value.

\item 
Apart from above presented case of the oscillatory instability
it is likely that there are similar instability in which other 
mechanical modes are involved (especialy violin ones which also have 
eigen frequencies several tens kHz and higher). There are also
additional instability for the pendelum mode in the mirror's 
suspension (in the case of small detuning of pumping optical frequency
out of resonance). These potencial "dangers" also deserve accurate analysis.

\end{enumerate}

We think that the parametric oscillatory instability effect can be
excluded in the laser gravitational antennae after this detailed
investigation.

\section*{Acknowledgements}

Authors are very grateful to H.~J.~Kimble, S.~Witcomb  and especially
to F.~Ya.~Khalili for help,
stimulating discussions and advises. This work was supported in part
by NSF and Caltech grants and by Russian Ministry of Industry and Science
and Russian Foundation of Basic Researches.

\appendix

\section{Lagrangian Approach}\label{app1}

Let us denote $q_0(t)$ and $q_1(t)$ as generalized coordinates for
the FP resonator optical modes with frequencies $\omega_0$ and
$\omega_1$ correspondingly, so that their vector potentials
($A_0,\ A_1$), electrical ($E_0,\ E_1$) and magnetic ($H_0,\ H_1$)
fields are the following:
\begin{eqnarray*}
A_i(t) &=& \sqrt{\frac{2\pi\, c^2}{ S_iL}}
    \left(f_i e^{ik_i z} - f_i^*e^{-ik_i z}\right)\, q_i(t),\\
E_i(t) &=& -\sqrt{\frac{2\pi  }{ S_iL}}
    \left(f_ie^{ik_i z} - f_i^*e^{-ik_i z}\right)\, \partial_t q_i(t),\\
H_i(t) &=& \sqrt{\frac{2\pi }{ S_iL}}
    \left(f_ie^{ik_i z} + f_i^*e^{-ik_i z}\right)\, \omega_i q_i(t),\\
f_i&=&f_i(\vec r_\bot,z), \quad S_i=\int |f_i|^2\vec dr_\bot.
\end{eqnarray*}
Let also denote $x(t)$ as generalized coordinate of the considered
elastic oscillations mode with displacement spatial distribution
described by the vector $\vec u(\vec r)$. Now we can write down
the lagrangian:
\begin{eqnarray*}
{\cal L} &=& {\cal L}_0 +{\cal L}_1 +{\cal L}_m+{\cal L}_{int},\\
{\cal L}_0 &=& \int \frac{L(\langle E_0\rangle^2 -\langle H_0\rangle^2)}{8\pi}
    \,\vec dr_\bot=
    \frac{\partial_t q_0^2}{2}-\frac{\omega_0^2q_0^2}{2},\\
{\cal L}_1 &=& \frac{\partial_t q_1^2}{2}-\frac{\omega_0^2q_1^2}{2},\\
{\cal L}_m &=& \frac{M(\partial_t x)^2}{2}- \frac{M\omega_m^2 x^2}{2},
    \\
M &=&\rho\int_V |\vec u(\vec r)|^2\, dV,\\
{\cal L}_{int} &=& -\int\left.\frac{x\, u_{z}
    \langle H_0+H_1\rangle^2}{8\pi}
    \right|_{z=0} d\vec r_\bot=\\
& = & -2\omega_0\omega_1\,  q_0q_1\, B\,\frac{x}{L},\\
B &=& \frac{\int f_0(\vec r_\bot) f_1(\vec r_\bot) u_{z} d\vec r_\bot}{
    \sqrt{\int |f_0|^2 d\vec r_\bot \int |f_1|^2 d\vec r_\bot } }
\end{eqnarray*}
We consider only one mechanical mode below. Now we can write down
the equations of motion (adding losses in each degree of freedom):
\begin{eqnarray*}
\partial_t^2 q_0 +2\delta_0 \partial_t q_0 +\omega_0^2 q_0 &=&
    - B \frac{2x}{L}\,\omega_0 \omega_1  q_1,\\
\partial_t^2 q_1 +2\delta_1 \partial_t q_1 +\omega_1^2 q_1 &=&
    - B\,\frac{2x}{L}\, \omega_1 \omega_0 \, q_0,\\
\partial_t^2 x +2\delta_m \partial_t x +\omega_m^2 x &=&
    - B \frac{2\omega_0\omega_1}{ML}\, q_0\, q_1.
\end{eqnarray*}
Introducing slowly varying amplitudes we can rewrite these
equations as:
\begin{eqnarray}
q_0(t) &=& D_0(t)\,e^{-i\omega_0t}+ D_0^*(t)\, e^{i\omega_0t},\nonumber\\
q_1(t) &=& D_1(t)\,e^{-i\omega_1t}+ D_1^*(t)\, e^{i\omega_1t},\nonumber\\
x(t)&=& X(t)\,e^{-i\omega_m t}+ X^*(t)\, e^{i\omega_mt},\nonumber\\
\Delta\omega &=& \omega_0-\omega_1-\omega_m,\nonumber\\
\partial_t D_0 +\delta_0 D_0  &=&
    \frac{i B X\,D_1 \omega_1 }
        {L }\, e^{i\Delta\omega t} ,
        \nonumber\\
\label{QQ1}
\partial_t D_1 +\delta_1 D_1  &=&
    \frac{iB X^*\,D_0 \omega_0 }
        {L }\, e^{-i\Delta\omega t},\\
\label{XX}
\partial_t X +\delta_m X &=&
    \frac{i\,B D_0D_1^*\omega_0\omega_1}{M\omega_m L }\,
    \,e^{-i\Delta\omega t},
\end{eqnarray}
We can see that this system (\ref{XX}, \ref{QQ1}) coincides with
(\ref{X}, \ref{Q1}) if $B=1$, and $M=m$.

For the simplest resonance case $\delta \omega=0$ it is easy to
substitute (\ref{XX}) into (\ref{QQ1}) and to obtain the condition
of parametric instability (in the frequency domain):
\begin{eqnarray*}
D_1(\delta_1- i\Omega) &=& \frac{iB D_0 \omega_0  }{L}\times
    \frac{-i\,B  D_0^*D_1\omega_0\omega_1}
        {M\omega_m L(\delta_m+i\Omega) },\\
&&\mbox{Condition of instability:}\\
1&<& \frac{B^2 |D_0|^2\omega_0^2\omega_1}{M\omega_m L^2  \delta_1\delta_m}
\end{eqnarray*}
Now we can express the energy ${\cal E}_0$ in mode "0" in terms of
$|D_0|^2$:
\begin{eqnarray*}
{\cal E}_0 &=&\frac{\partial_t q_0^2}{2}+
    \frac{\omega_0^2q_0^2}{2}=\\
&=& \frac{1}{2}\left( (-i\omega_0)^2
    \left[D_0\,e^{-i\omega_0t}- D_0^*\, e^{i\omega_0t}\right]^2+\right.\\
&&\qquad +\left.\omega_0^2
    \left[D_0\,e^{-i\omega_0t}+ D_0^*\, e^{i\omega_0t}\right]^2\right)=\\
&=& 2\omega_0^2|D_0|^2.
\end{eqnarray*}
Now we can write down the condition of parametric instability in
the following form:
\begin{eqnarray}
&&\frac{{\cal E}_0 B }{2m\omega_m^2 L^2}\times
    \frac{\omega_1\omega_m}{\delta_1\delta_m}>1,\\
\Lambda &=& \frac{B^2 m}{M}=
    \frac{V\left(\int f_0(\vec r_\bot) f_1(\vec r_\bot)
        u_{z} d\vec r_\bot\right)^2}{
    \int |f_0|^2 d\vec r_\bot \int |f_1|^2 d\vec r_\bot
        \int |\vec u|^2 d V }
\end{eqnarray}
which accurately coincides with (\ref{CIL}, \ref{Lambda}).

Let us deduce the instability condition for nonresonance case. We
are looking for the solution of (\ref{Q1}, \ref{X}) in the
following form:
\begin{eqnarray*}
D_1(t) &=& D_1 e^{\lambda_- t},\quad
    X^*(t) = X^* e^{\lambda_+ t},\\
\lambda_- = \lambda - {i\Delta \omega \over 2},&&
\lambda_+ = \lambda + {i\Delta \omega\over 2},
\end{eqnarray*}
and writing down the characteristic equation as:
\begin{eqnarray*}
(\lambda_+ +\delta_1)(\lambda_- +\delta_m)-A&=&0,\quad
\frac{D_0^2\omega_0^2\omega_1\Lambda}{m \omega_m L^2}=A.
\end{eqnarray*}
The solutions of characteristic equation are:
\begin{eqnarray*}
\lambda_{1,2} &=& -\frac{\delta_1+\delta_m}{2}\pm \sqrt{\rm Det},\\
{\rm Det} &=&
     \left(\frac{\delta_1 -\delta_m}{2} -\frac{i\Delta\omega}{2} \right)^2
    +A.
\end{eqnarray*}
The condition of instability is the following:
\begin{equation}
\label{CIA}
\Re \sqrt{\rm Det} > \frac{\delta_1+\delta_m}{2}.
\end{equation}
Using a convenient formula:
\begin{eqnarray*}
{\rm Det} &=& a+ib,\\
\Re\sqrt{\rm Det} &=&\frac{\sqrt 2}{2} \sqrt{\sqrt{a^2+b^2}+a}.
\end{eqnarray*}
we can rewrite the condition (\ref{CIA}) as:
\begin{eqnarray}
\label{eq36}
\frac{1}{2} \left(\sqrt{a^2+b^2}+a\right) &> &\left(
        \frac{\delta_1+\delta_m}{2}\right)^2,\\
a^2+b^2 &=& A^2+ \left(\frac{(\delta_1-\delta_m)^2}{4} +
    \frac{\Delta \omega^2}{4}\right)^2+\nonumber\\
&&+2A\left(\frac{(\delta_1-\delta_m)^2}{4} -
    \frac{\Delta \omega^2}{4}\right)
\end{eqnarray}
Note that for the resonance case ($\Delta \omega =0$) the solution
of (\ref{CIA} or \ref{eq36}) is known: $A> \delta_1\delta_m$. For
our case $\delta_m \ll \delta_1$ it means that $A\ll \delta_1^2$.
Therefore for small detuning $\Delta\omega\ll \delta_1$ we can
expand $a^2+b^2$ in series in terms of $A$ and rewrite condition
(\ref{eq36}) as:
\begin{eqnarray}
\frac{1}{2} \left(\sqrt{a^2+b^2}+a\right) &\simeq&
    \frac{A}{2}+\frac{(\delta_1-\delta_m)^2}{4} + \nonumber\\
&&+ \frac{A}{2}\,\frac{
    \frac{(\delta_1-\delta_m)^2}{4} -
    \frac{\Delta \omega^2}{4}}
    {\frac{(\delta_1-\delta_m)^2}{4} +
    \frac{\Delta \omega^2}{4}
    }\nonumber\\
A\frac{
    \frac{(\delta_1-\delta_m)^2}{4} }
    {\frac{(\delta_1-\delta_m)^2}{4} +
    \frac{\Delta \omega^2}{4}
    }   &> & \delta_1\delta_m,\nonumber\\
\label{smallo}
\mbox{ Or}\quad  A > \delta_1\delta_m &\times &
    \frac{(\delta_1-\delta_m)^2+\Delta \omega^2}{(\delta_1-\delta_m)^2}.
\end{eqnarray}
Let us underline that condition (\ref{smallo}) is obtained for
small detuning $\Delta\omega \ll \delta_1$. However, considering
situation more attentively one can conclude that expansion in
series (and consequently the formula (\ref{smallo}) ) is valid for
the condition:
\begin{eqnarray}
A&\ll & \frac{(\delta_1-\delta_m)^2}{4} +
    \frac{\Delta \omega^2}{4}
\end{eqnarray}
We see that this condition is fulfilled for the solution (\ref{smallo}).
Therefore we conclude that solution (\ref{smallo}) is approximately valid
{\em for any detunings} $\Delta \omega$.

\end{document}